\documentclass[12pt]{article}
\usepackage{stmaryrd}
\usepackage{amssymb}
\usepackage{graphics}
\usepackage{epsfig}
\usepackage{a4wide}

\begin{document}

\title{ Prediction feedback in intelligent traffic systems\footnote{Supported by National Natural Science Foundation
of China.}}
\author{Dong Chuan-Fei, Ma Xu, Wang Guan-Wen, Sun Xiao-Yan, and Wang Bing-Hong\footnote{bhwang@ustc.edu.cn} \\
{\small Department of Modern Physics and Nonlinear Science Center, University of Science and Technology}\\
{\small of China (USTC), Hefei, Anhui 230026, P.R.China}  }

\date{}
\maketitle \vskip 15mm
\begin{abstract}
The optimal information feedback has a significant effect on many
socioeconomic systems like stock market and traffic systems aiming
to make full use of resources. In this paper, we studied dynamics of
traffic flow with real-time information provided and the influence
of a feedback strategy named prediction feedback strategy is
introduced, based on a two-route scenario in which dynamic
information can be generated and displayed on the board to guide
road users to make a choice. Our model incorporates the effects of
adaptability into the cellular automaton models of traffic flow and
simulation results adopting this optimal information feedback
strategy have demonstrated high efficiency in controlling spatial
distribution of traffic patterns compared with the other three
information feedback strategies, i.e., vehicle number and flux.

\end{abstract}

{\large\bf PACS: 12.38.Bx, 12.15.Lk, 14.70.Hp, 14.65.Ha }

\vfill \eject

\baselineskip=0.32in

\renewcommand{\theequation}{\arabic{section}.\arabic{equation}}
\renewcommand{\thesection}{\Roman{section}.}
\newcommand{\nb}{\nonumber}

\newcommand{\Dir}{\kern -6.4pt\Big{/}}
\newcommand{\Dirin}{\kern -10.4pt\Big{/}\kern 4.4pt}
\newcommand{\DDir}{\kern -7.6pt\Big{/}}
\newcommand{\DGir}{\kern -6.0pt\Big{/}}

\makeatletter      
\@addtoreset{equation}{section}
\makeatother       

\section{Introduction}
\par
Vehicular traffic flow and related problems have triggered great
interests of a community of physicists in recent years because of
its various complex behaviors.\cite{s1, s2, s3} and also a lot of
theories have been proposed such as car-following theory\cite{s4},
kinetic theory\cite{s5, s6, s7, s8, s9, s10, s11} and
particle-hopping theory\cite{s12, s13}.These theories have the
advantage of alleviating the traffic congestion and enhance the
capacity of existing infrastructure. Although dynamics of traffic
flow with real-time traffic information have been extensively
investigated\cite{s14, s15, s16, s17, s18, s19}, finding a more
efficient feedback strategy is an overall task. Recently, some
real-time feedback strategies have been put forward, such as Travel
Time Feedback Strategy(TTFS)\cite{s14, s20} and Mean Velocity
Feedback Strategy(MVFS)\cite{s14,s21}and Congestion Coefficient
Feedback Strategy(CCFS)\cite{s14,s22}. It has been proved that MVFS
is more efficient than that of TTFS which brings a lag effect to
make it impossible to provide the road users with the real situation
of each route\cite{s21} and CCFS is more efficient than that of MVFS
because of the fact that the random brake mechanism of the
Nagel-Schreckenberg(NS) model\cite{s12} brings fragile stability of
velocity\cite{s22}. However, CCFS is still not the best one due to
the fact that its feedback is not in time, so it cannot reflect the
road situation immediately and some other reasons which will be
discussed delicately in this paper. In order to provide road users
with better guidance, a strategy named prediction feedback strategy
(PFS) is presented. We report the simulation results adopting four
different feedback strategies in a two-route scenario with single
route following the NS mechanism.

\par
The paper is arranged as following: In Sec. II the NS model and
two-route scenario are briefly introduce, together with four
feedback strategies of TTFS, MVFS, CCFS and PFS all depicted in more
detail. In Sec. III some simulation results will be presented and
discussed based on the comparison of four different feedback
strategies. The last section will make some conclusions.

\vskip 10mm
\section{THE MODEL AND FEEDBACK STRATEGIES}
\par
\textbf{A. NS mechanism}

\par
The Nagel-Schreckenberg (NS) model is so far the most popular and
simplest cellular automaton model in analyzing the traffic
flow\cite{s1,s2,s3,s12,s23}, where the one-dimension CA with
periodic boundary conditions is used to investigate highway and
urban traffic. This model can reproduce the basic features of real
traffic like stop-and-go wave, phantom jams, and the phase
transition on a fundamental diagram. In this section, the NS
mechanism will be briefly introduced as a base of analysis.

\par
The road is subdivided into cells with a length of $\Delta$x=7.5 m.
Let \emph{N} be the total number of vehicles on a single route of
length \emph{L}, then the vehicle density is
$\rho$=\emph{N}/\emph{L}. $\emph{g}_{n}$(t) is defined to be the
number of empty sites in front of the \emph{n}th vehicle at time
\emph{t}, and $\emph{v}_{n}$(t) to be the speed of the \emph{n}th
vehicle, i.e., the number of sites that the \emph{n}th vehicle moves
during the time step \emph{t}. In the NS model, the maximum speed is
fixed to be $\emph{v}_{max}$=\emph{M}. In the present paper, we set
\emph{M}=3 for simplicity.

\par
The NS mechanism can be decomposed to the following four rules
(parallel dynamics):

\par
Rule 1. Acceleration: $\emph{v}_{i} \leftarrow
 min(\emph{v}_{i}+1,M)$;

\par
Rule 2. Deceleration: $\emph{v}_{i}^{'} \leftarrow
 min(\emph{v}_{i},\emph{g}_{i})$;

\par
Rule 3. Random brake: with a certain brake probability \emph{P} do
$\emph{v}_{i}^{''} \leftarrow max(\emph{v}_{i}^{'}-1,0)$; and

\par
Rule 4. Movement: $\emph{x}_{i} \leftarrow
\emph{x}_{i}+\emph{v}_{i}^{''}$;

\par
The fundamental diagram characterizes the basic properties of the NS
model which has two regimes called "free-flow" phase and "jammed"
phase. The critical density, basically depending on the random brake
probability \emph{p}, divides the fundamental diagram to these two
phases.

\par
\textbf{B. Two-route scenario}

\par
Wahle \emph{et al}.\cite{s20} first investigated the two-route model
in which road users choose one of the two routes according to the
real-time information feedback. In the two-route scenario, it is
supposed that there are two routes A and B of the same length
\emph{L}. At every time step, a new vehicle is generated at the
entrance of two routes and will choose one route. If a vehicle
enters one of two routes, the motion of it will follow the dynamics
of the NS model. As a remark, if a new vehicle is not able to enter
the desired route, it will be deleted. The vehicle will be removed
after it reaches the end point.

\par
Additionally, two types of vehicles are introduced: dynamic and
static vehicles. If a driver is a so-called dynamic one, he will
make a choice on the basis of the information feedback \cite{s20},
while a static one just enters a route at random ignoring any
advice. The density of dynamic and static travelers are
$\emph{S}_{dyn}$ and $1-\emph{S}_{dyn}$, respectively.

\par
The simulations are performed by the following steps: first, set the
routes and board empty; then, after the vehicles enter the routes,
according to four different feedback strategies, information will be
generated, transmitted, and displayed on the board at every time
step. Then the dynamic road users will choose the route with better
condition according to the dynamic information at the entrance of
two routes.

\par
\textbf{C. Related definitions}

\par
The roads conditions can be characterized by flux of two routes, and
flux is defined as follows:
\begin{eqnarray}
\emph{F}=V_{mean}\rho=V_{mean}\frac{N}{L}
\end{eqnarray}
where $V_{mean}$ represents the mean velocity of all the vehicles on
one of the roads, \emph{N} denotes the vehicle number on each road,
and \emph{L} is the length of two routes. Then we describe four
different feedback strategies, respectively.

\par
TTFS: At the beginning, both routes are empty and the information of
travel time on the board is set to be the same. Each driver will
record the time when he enters one of the routes. Once a vehicle
leaves the two-route system, it will transmit its travel time on the
board and at that time a new dynamic driver will choose the road
with shorter time.

\par
MVFS: Every time step, each vehicle on the routes transmits its
velocity to the traffic control center which will deal with the
information and display the mean velocity of vehicles on each route
on the board. Road users at the entrance will choose one road with
larger mean velocity.

\par
CCFS: Every time step, each vehicle transmits its signal to
satellite, then the navigation system (GPS) will handle that
information and calculate the position of each vehicle which will be
transmitted to the traffic control center. The work of the traffic
control center is to compute the congestion coefficient of each road
and display it on the board. Road users at the entrance will choose
one road with smaller congestion coefficient.

\par
The congestion coefficient is defined as
\begin{equation}
C=\sum_{i=1}^{m} n_i^{w}.
\end{equation}
Here, $\emph{n}_{i}$ stands for vehicle number of the \emph{i}th
congestion cluster in which cars are close to each other without a
gap between any two of them. Every cluster is evaluated a weight
\emph{w}, here \emph{w}=2\cite{s22}.

\par
PFS: We do the work about PFS on the basis of CCFS, because CCFS is
the best one among the three strategies above.

\par
Every time step, the traffic control center will receive data from
the navigation system (GPS) like CCFS, and the work of the center is
to compute the congestion coefficient of each road and simulate the
road situation in the future making use of the current road
situation by using CCFS. Then display it on the board. Road users at
the entrance will choose one road with smaller congestion
coefficient. For example, if the prediction time($\emph{T}_{p}$) is
50 seconds and the current time is 100th second, the traffic control
center will simulate the road situation at the next 50 seconds by
using CCFS and predict the road situation at 150th second, then show
the result on the board at the entrance of the road. Finally the
road users at 100th second will choose one road with smaller
congestion coefficient at 150th second predicted by the new
strategy. So as to analogize, the road user at the entrance at 101th
second will choose one road with small congestion coefficient at
151th second predicted by the new strategy like explained above and
so on.

\par
Compared with the former work\cite{s20,s21,s22}, another important
difference we have done in this paper is that we set the two-route
system has only one entrance and one exit as it shows in the Fig.1
while the two-route system before has one entrance and two exits. So
we do research work based on the two-route system which is more
close to the reality instead of simply repeating other work. The
rules at the exit of the two-route system are as following:
\begin{figure}[htbp]
\vspace*{-0.3cm} \centering
\includegraphics*{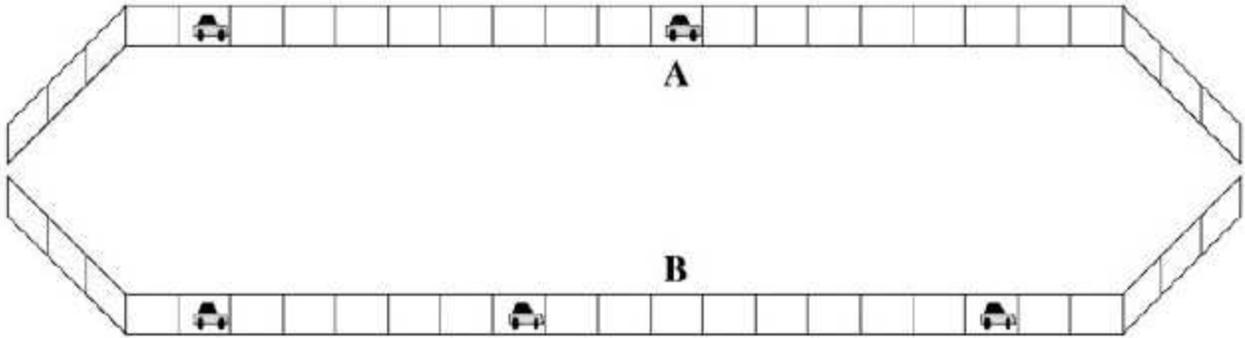}
\vspace*{-0.3cm} \centering \caption{\label{fig1} The two-route
system only has one entrance and one exit which is different from
the road situation in former work.}
\end{figure}

\par
(a) At the end of two routes, the car that is nearer to the exit
goes first.

\par
(b) If the cars at the end of two routes have the same distance to
the exit, which one drives faster, which one goes out first.

\par
(c) If the cars at the end of two routes have the same distance to
the exit and speed, the car in the route which has more cars goes
first.

\par
(d) If the rule (a), (b) and (c) are satisfied at the same time,
then the cars go out randomly.

\par
In the following section, performance by using four different
feedback strategies will be shown and discussed in more detail.

\vskip 10mm
\section{SIMULATION RESULTS}

\par
All simulation results shown here are obtained by 30000 iterations
excluding the initial 5000 time steps. Figure 2 shows the dependence
of average flux and prediction time($\emph{T}_{p}$) by using the new
strategy. As to the routes' processing capacity. We can see that in
Fig.2 there are positive peak structures at the vicinity of
$\emph{T}_{p}$ $\sim$ 60. So we will use $\emph{T}_{p}$=60 in the
following paragraphs.
\begin{figure}
\centering
\includegraphics[scale=1.0]{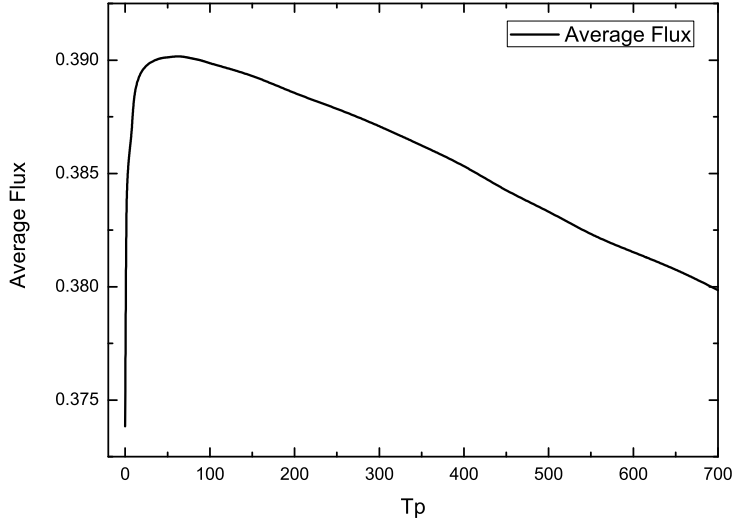}
\caption{\label{fig2} Average flux by performing different
prediction time($\emph{T}_{p}$). The parameters are \emph{L}=2000,
\emph{p}=0.25, and $\emph{S}_{dyn}$=0.5.}
\end{figure}

\par
In contrast with PFS, the flux of two routes adopting CCFS, MVFS and
TTFS shows oscillation obviously (see Fig.3) due to the information
lag effect\cite{s22}. This lag effect can be understood as that the
other three strategies cannot reflect the road current situation.
Another reason for the oscillation is that two-route system only has
one exit, therefore, only one car can go out at one time step which
may result in the traffic jam to happen at the end of the routes and
the new strategy can predict the effects to the route situation
caused by the traffic jam at the end of the route, therefore, the
new strategy may improve the road situation. Compared to CCFS, the
performance adopting PFS is remarkably improved, not only on the
value but also the stability of the flux. Therefore as to the flux
of the two-route system, PFS is the best one.
\begin{figure}
\centering
\includegraphics[scale=0.7]{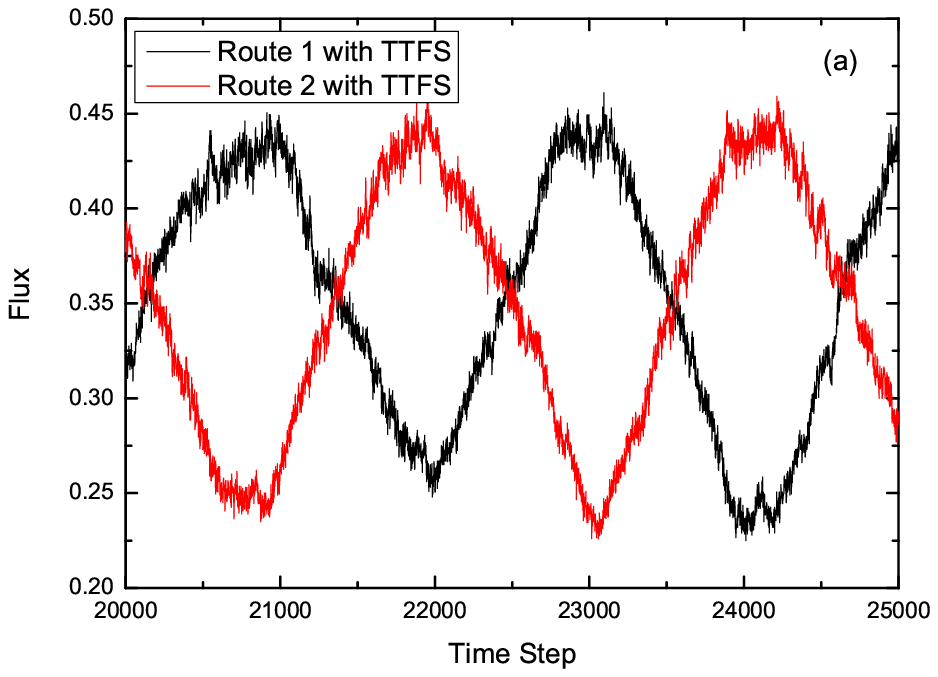}
\includegraphics[scale=0.7]{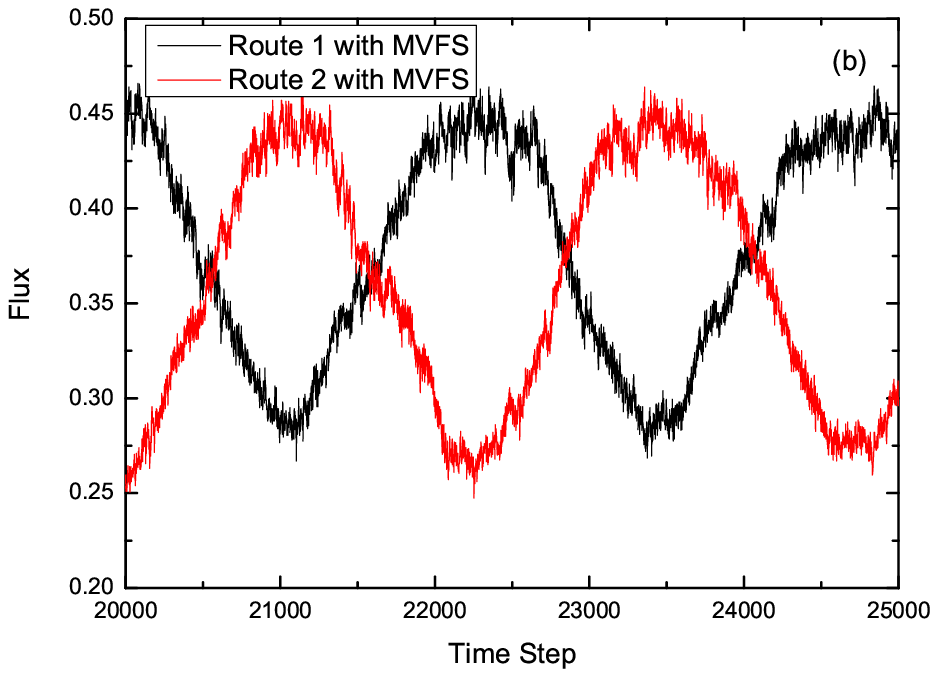}
\includegraphics[scale=0.7]{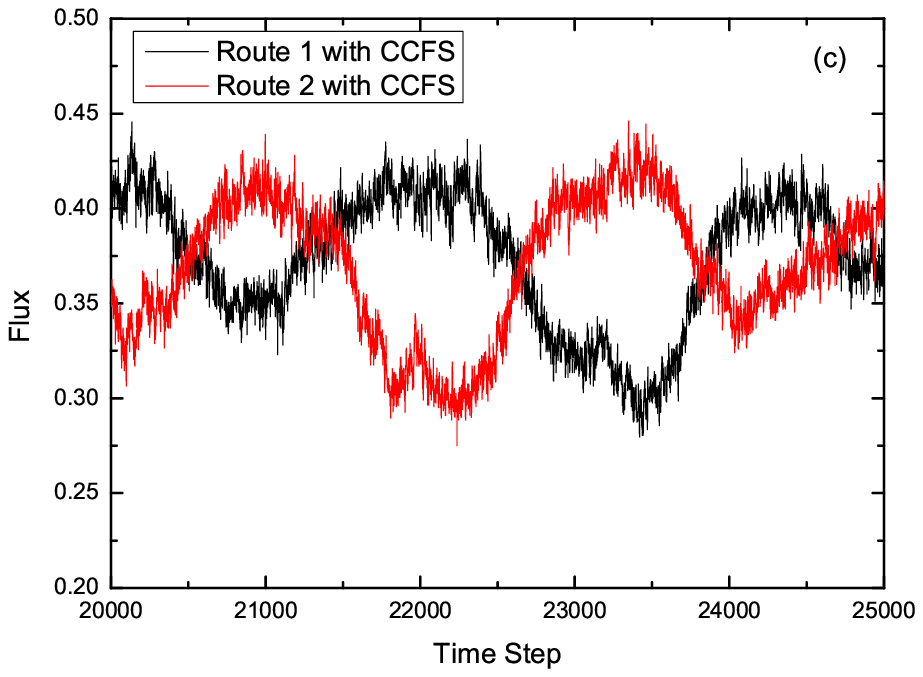}
\includegraphics[scale=0.7]{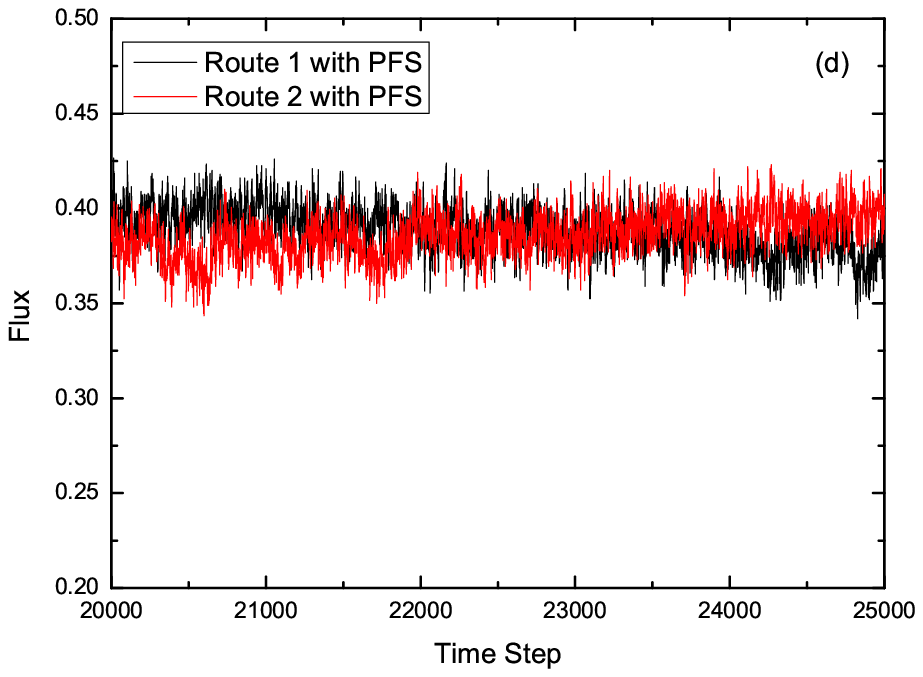}
\caption{\label{fig3} (Color online)(a) Flux of each route with
travel time feedback strategy. (b) Flux of each route with mean
velocity feedback strategy. (c) Flux of each route with congestion
coefficient feedback strategy. (d) Flux of each route with
prediction feedback strategy. The parameters are \emph{L}=2000,
\emph{p}=0.25, $\emph{S}_{dyn}$=0.5, and $\emph{T}_{p}$=60. }
\end{figure}

\par
In Fig.4, vehicle number versus time step shows almost the same
tendency as Fig.3, the routes' accommodating capacity is greatly
enhanced with an increase in vehicle number from 290 to 780, so
perhaps the high flux of two routes with PFS are mainly due to the
increase of vehicle number. Maybe someone will ask why the vehicle
number in Fig.4 using other three strategies is larger than the
figures show in the former work\cite{s22}. The reason is that the
road situation is different from the work before. The two-route
system in this paper only has one exit, therefore, only one car can
go out at one time step which will lead to the increasing of vehicle
number in each route.
\begin{figure}
\centering
\includegraphics[scale=0.7]{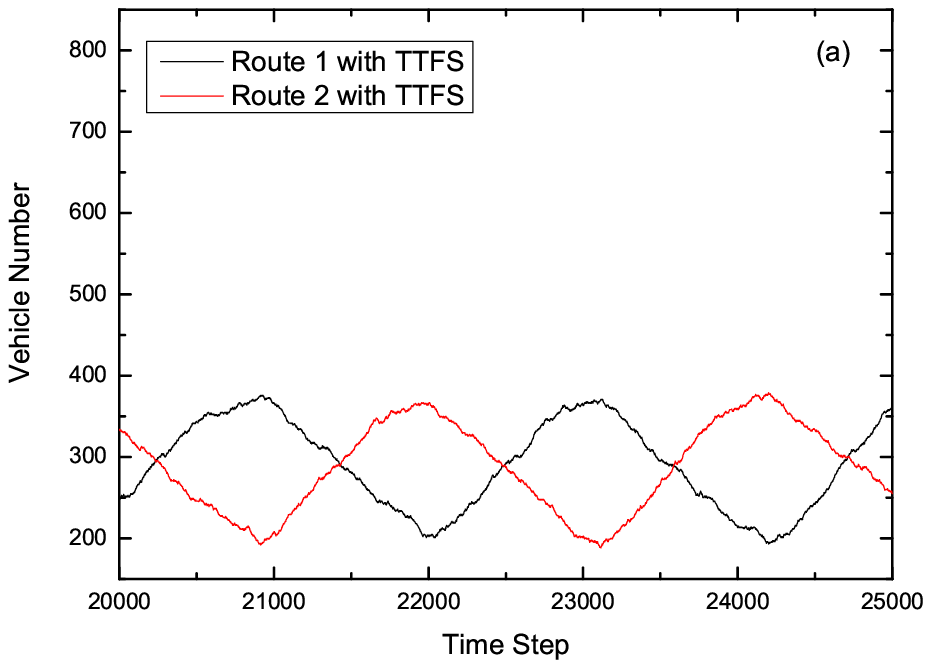}
\includegraphics[scale=0.7]{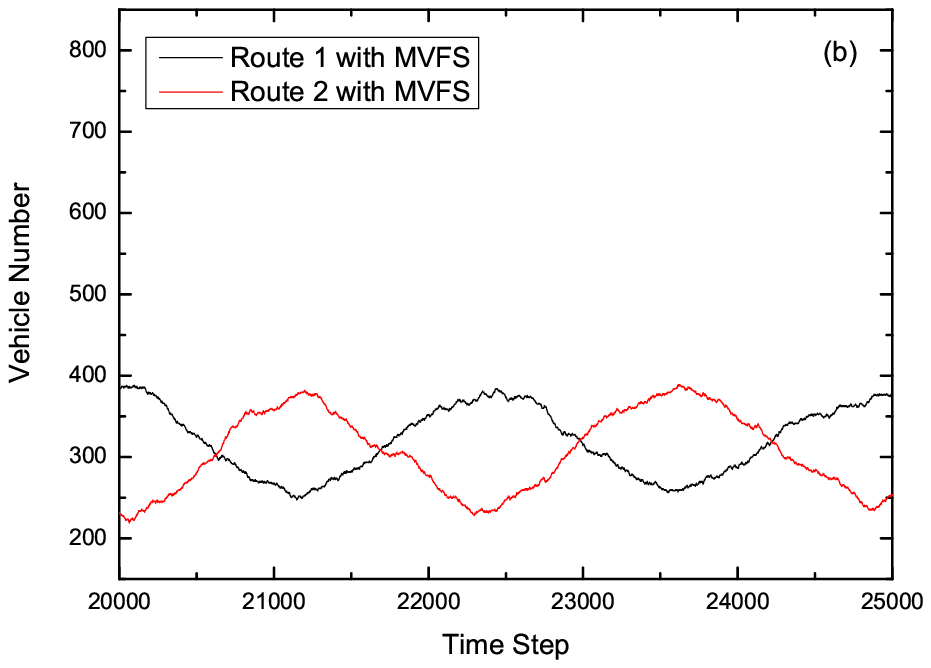}
\includegraphics[scale=0.7]{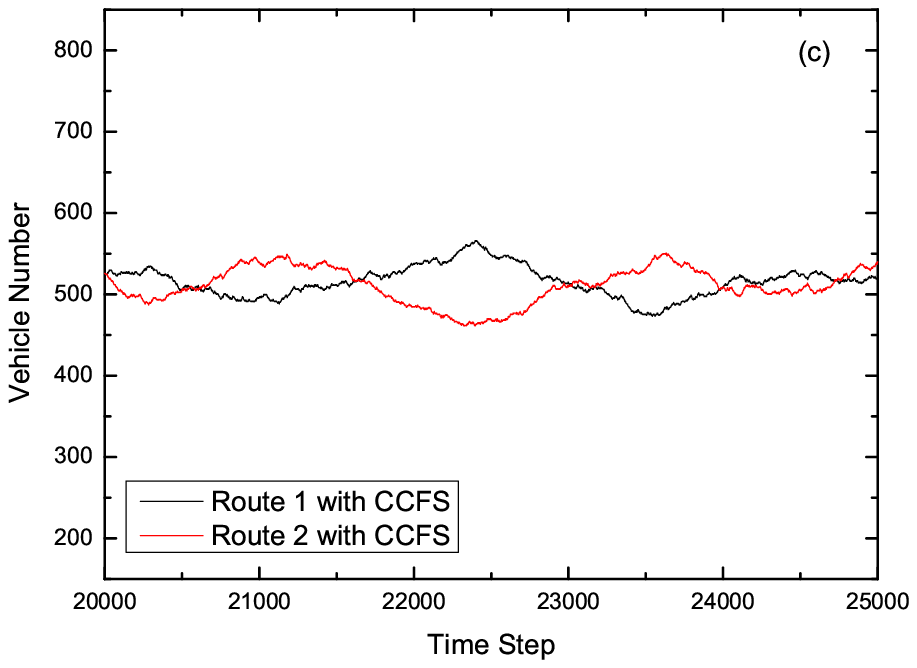}
\includegraphics[scale=0.7]{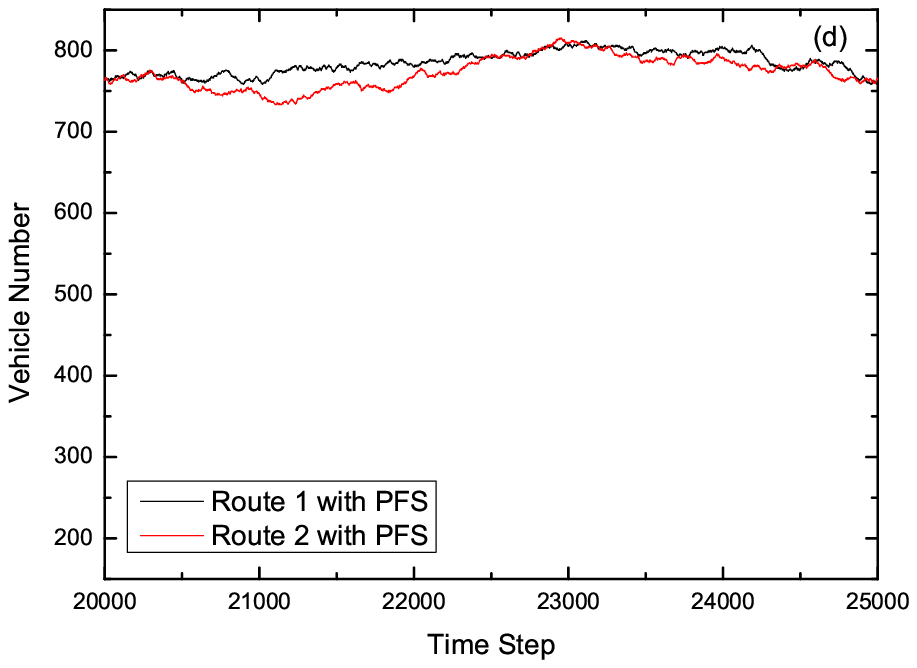}
\caption{\label{fig4} (Color online)(a) Vehicle number of each route
with travel time feedback strategy. (b) Vehicle number of each route
with mean velocity feedback strategy. (c) Vehicle number of each
route with congestion coefficient feedback strategy. (d) Vehicle
number of each route with prediction feedback strategy. The
parameters are set the same as in Figure 3.}
\end{figure}

\par
In Fig.5, speed versus time step shows that although the speed is
stablest by using the new strategy, it is the lowest among the four
different strategies. The reason is that the routes' accommodating
capacity is best by using the new strategy and as mentioned above
the road has only one exit and only one car can go out at one time
step, therefore, the more cars, the lower speed. Fortunately, flux
consists of two parts, mean velocity and vehicle density, therefore,
as long as the vehicle number (because the vehicle density is
$\rho$=\emph{N}/\emph{L}, and the \emph{L} is fixed to be 2000, so
$\rho$ $\propto$ vehicle number (\emph{N})) is large enough, the
flux can also be the largest.
\begin{figure}
\centering
\includegraphics[scale=0.675]{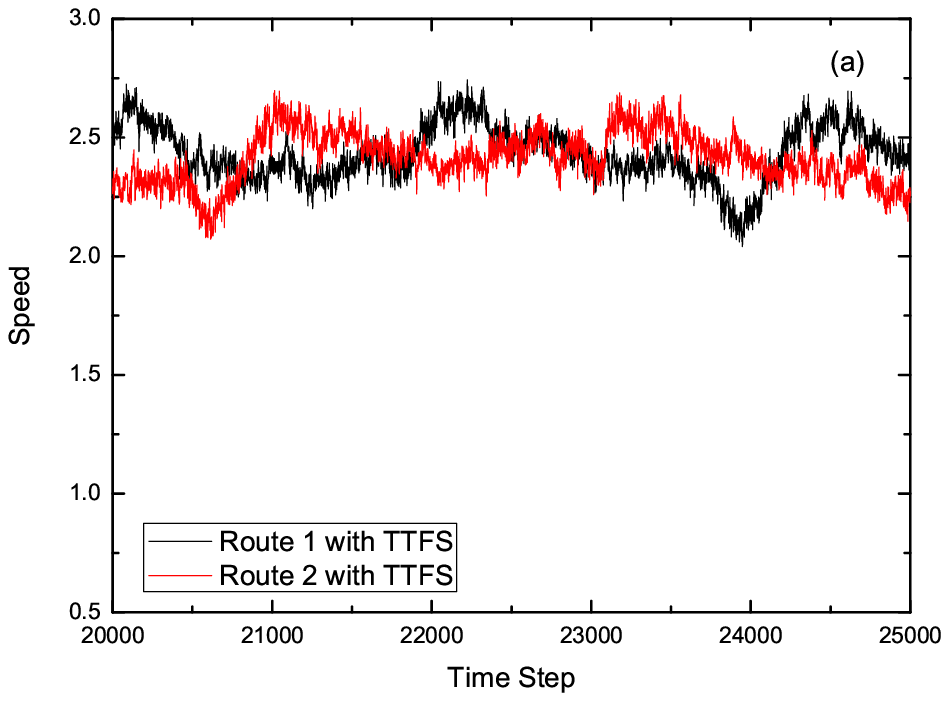}
\includegraphics[scale=0.7]{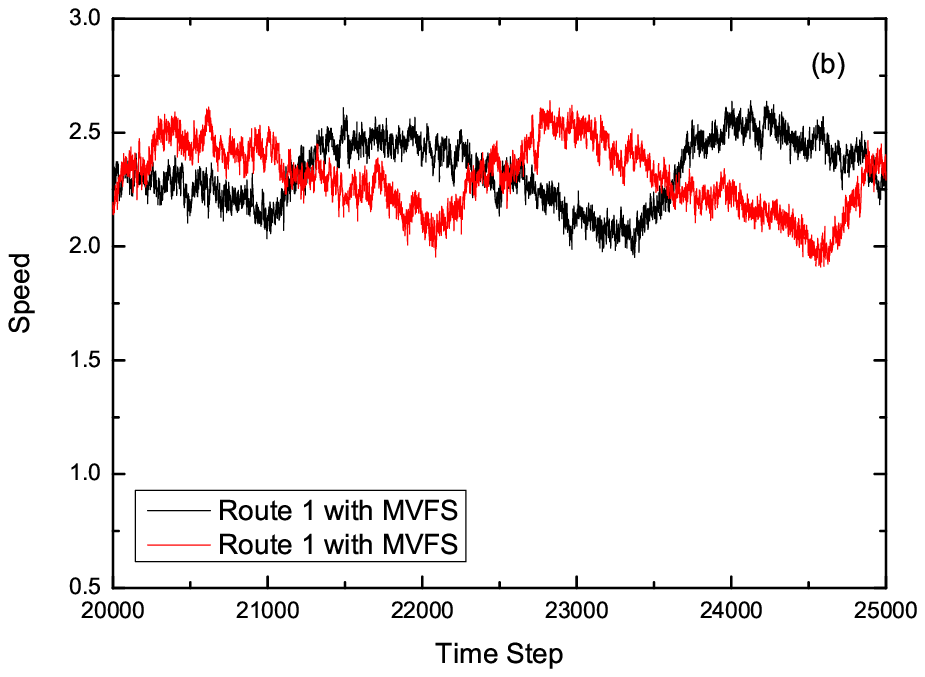}
\includegraphics[scale=0.7]{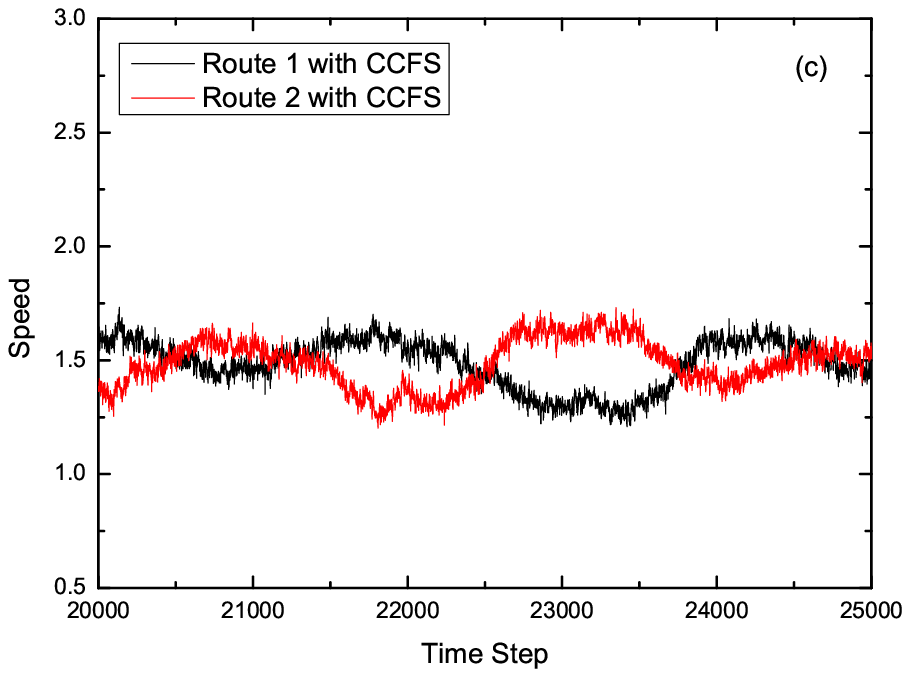}
\includegraphics[scale=0.7]{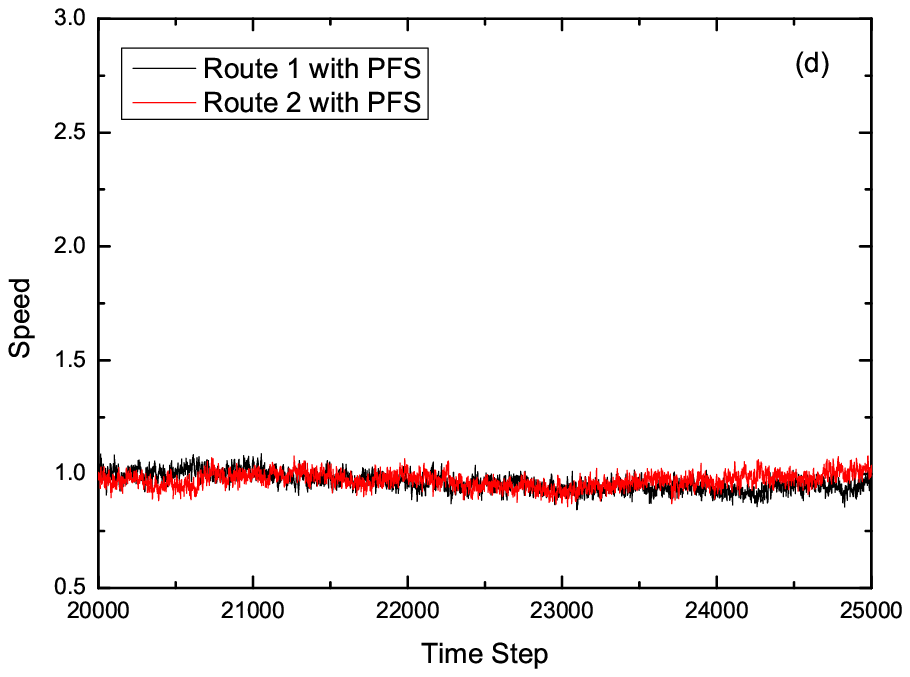}
\caption{\label{fig5} (Color online)(a) Average speed of each route
with travel time feedback strategy. (b) Average speed of each route
with mean velocity feedback strategy. (c) Average speed of each
route with congestion coefficient feedback strategy. (d) Average
speed of each route with prediction feedback strategy. The
parameters are set the same as in Figure 3.}
\end{figure}

\par
Fig.6 shows that the average flux fluctuates feebly with a
persisting increase of dynamic travelers by using the new strategy.
As to the routes' processing capacity, the new strategy is proved to
be the most proper one because the flux is always the largest at
each $\emph{S}_{dyn}$ value and even increases with a persisting
increase of dynamic travelers.
\begin{figure}
\centering
\includegraphics[scale=1.2]{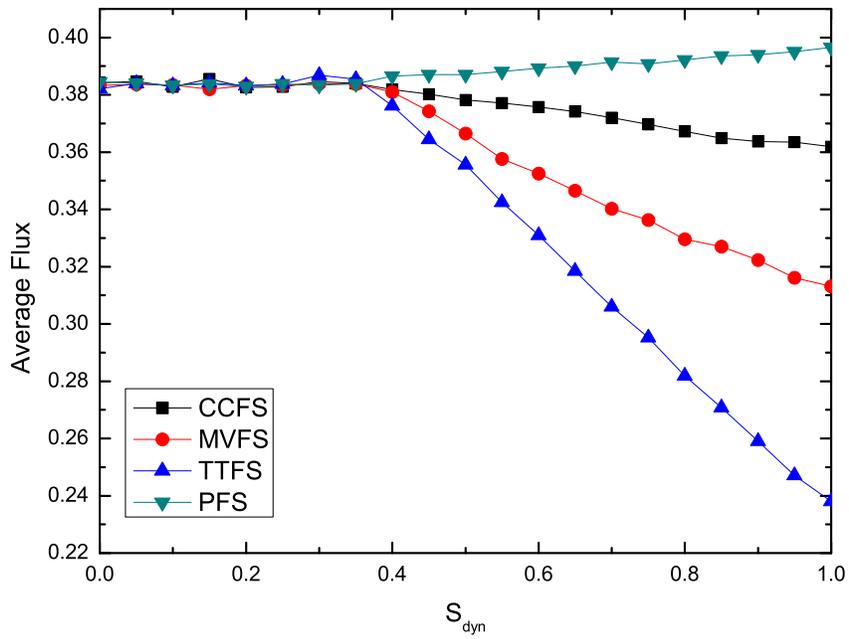}
\caption{\label{fig6} (Color online) Average flux by performing
different strategy vs $\emph{S}_{dyn}$; \emph{L} is fixed to be
2000, and $\emph{T}_{p}$ is fixed to be 60.}
\end{figure}

\vskip 10mm
\section{CONCLUSION}
\par
We obtain the simulation results of applying four different feedback
strategies, i.e., TTFS, MVFS, CCFS and PFS on a two-route scenario
all with respect to flux, number of cars, speed, average flux versus
$\emph{T}_{p}$ and average flux versus $\emph{S}_{dyn}$. The results
indicates that the PFS strategy has more advantages than the three
former ones in the two-route system which has only one entrance and
one exit. The highlight of this paper is that it brings forward a
new quantity namely prediction time ($\emph{T}_{p}$) to radically
improve road conditions. In contrast with the three old strategies,
the PFS strategy can bring a significant improvement to the road
conditions, including increasing vehicle number and flux, reducing
oscillation, and that average flux increases with increase of
$\emph{S}_{dyn}$. And it can be understood because the new strategy
can eliminate the lag effect. The numerical simulations demonstrate
that the prediction time($\emph{T}_{p}$) play an very important role
in improving the road situation.

\par
Due to the rapid development of modern scientific technology, it is
not difficult to realize PFS. If only a navigation system (GPS) is
installed in each vehicle, thus the position information of vehicles
will be known, then the PFS strategy can come true through
computational simulation by using the CCFS strategy and also it will
cost no more than CCFS because the computers using to compute the
congestion coefficient can also simulate the road situation in the
future. Taking into account the reasonable cost and more accurate
description of road conditions, we think that this strategy shall be
applicable.

\vskip 10mm
\par
\noindent{\large\bf Acknowledgments:} This work has been partially
supported by the National Basic Research Program of China (973
Program No. 2006CB705500), the National Natural Science Foundation
of China (Grant Nos. 60744003, 10635040,10532060), the Specialized
Research Fund for the Doctoral Program of Higher Education of China
(Grant No. 20060358065) and  National Science Fund for Fostering
Talents in Basic Science (J0630319).

\end{document}